Costantino Sigismondi
*Università di Roma "La Sapienza"*
sigismondi@icra.it

# L'Epistolario di Gerberto, Papa Astronomo

**Abstract:** A new Italian edition of the Letters of Gerbert, the astronomer, scientist and philosopher who become pope Silvester II, with their papal privileges has been published in the occasion of the international year of astronomy.
The italian translation (1980) of Maria Giulia Panvini Carciotto (1949-1996), has been co-edited by Costantino Sigismondi and Paolo Rossi. The volume, according to the will of the relatives of Dr. Panvini Carciotto is available either in press and as free e-book on the web indexed with the words "Gerberto Epistolario".

### Introduzione: Gerberto, dalle sue lettere all'uomo

Le lettere di Gerberto di Aurillac (ca. 945-1003) [1] sono delle importanti tessere per capire il complesso mosaico della storia del 10° secolo. Il grande docente della scuola cattedrale di Reims (dal 972 al 989), anche nella qualità di segretario dell'Arcivescovo Adalberone, e dal 980 anche Abate di Bobbio, scrisse queste lettere per vari destinatari: imperatori e imperatrici della dinastia degli Ottoni, re di Francia, regine e nobildonne, vescovi ed abati, monaci e scolastici. È documentato anche il suo tormentato vescovato a Reims (991-996), il suo ritiro in Sassonia presso la corte di Ottone III (996-998), prima di ottenere la diocesi di Ravenna (998-999) e poi il sommo pontificato (999-1003).

Lo stile è sempre essenziale: *non trasgrediamo, dicendo di più, le norme epistolari* [2, lettera 45], *ce la sbrighiamo in poche parole* [2, lettera 48]…e trovano spazio sia citazioni di classici latini come Cicerone e Virgilio, sia dalle Sacre Scritture o dalla *Regula* di S. Benedetto.

Le lettere scientifiche completano il vastissimo panorama culturale abbracciato da Gerberto.

Paolo Rossi –direttore del Dipartimento di Fisica dell'Università di Pisa e responsabile della collana Fonti Tradotte per l'Alto Medioevo– ha tradotto i prvilegi papali dall'edizione Inglese di Harriett Pratt-Lattin [3]: sono i documenti pontifici di Silvestro II e tratteggiano il coronamento della sua visione politica e religiosa.

Gerberto, da Papa, trovò quello che aveva preparato durante tutta la sua vita politica: dopo aver difeso strenuamente l'impero del giovanissimo Ottone III, il designato *Romanorum Christus* [4] (che era rimasto orfano già a 3 anni), poté estendere i confini giuridici della Chiesa fino in Polonia e in Ungheria (proprio nell'anno 1000), con la fondazione delle prime diocesi.

Gerberto vide nel suo pontificato la massima espansione della Chiesa Cattolica, ancora unita, nell'ecumène allora conosciuto. Lo scisma degli ortodossi avvenne nel 1054, ma Ottone III, figlio della greca Teòfania e Ottone II: *Greco per stirpe Romano per impero* [2, Lettera 187], incarnava e garantiva con la sua persona l'unità della Chiesa *una* e *santa*.

Dobbiamo aspettare l'età moderna perché la cattolicità torni ad espandersi sia numericamente che geograficamente oltre i limiti raggiunti nell'anno 1000.

### Lettere scientifiche

Sono numerose le lettere di argomento scientifico, indirizzate per lo più ai suoi studenti. Esse tradiscono la sua incessante attività di studio ed aggiornamento *nel riposo e negli affari insegniamo ciò che sappiamo e impariamo ciò che non sappiamo* [2, Lettera 44], attendendo *tempi migliori nei quali si possano risuscitare gli studi già da un pezzo in noi morti* quando le circostanze avverse gli toglievano tempo e serenità per gli studi, ed anche *per un pezzo di legno lavorato al tornio* [2, Lettera 152]. Così Gerberto fu il più importante docente della seconda metà del 10° secolo, capace di trasmettere, anche mediante repliche e creazioni ex-novo di strumenti (*al tornio*), primo nell'Europa cristiana, gli elementi di astronomia della cultura araba ed i numeri indo-arabi [4].

Il ruolo chiave di Gerberto [cfr. 5, 6] nella storia dell'astronomia è stato riconosciuto anche dal prof. Nha Il-Seong, presidente della commissione 43 di Storia dell'Astronomia dell'IAU "The establishments of these observatories may be referenced for many readers by two European highlights, the great scholar Pope Silvester II in the 11th century and Galileo and Kepler in the 17th century". [7]. Tuttavia la sua umiltà intellettuale ne garantisce l'autorità: *non volere mettere in pratica per mio consiglio quelle cose che pertengono ai medici, tenendo conto soprattutto del fatto che io ho soltanto preso contezza della loro scienza, ma ho sempre rifuggito dal sostituirmi ad essi* [2, Lettera 151]: per sua ammissione implicita già alla fine del 10° secolo non esistevano più gli esperti di tutto lo scibile umano.

### Le lettere di Gerberto come strumenti didattici

Questo epistolario, originariamente pubblicato nei Quaderni di Filologia Medievale della Facoltà di Magistero di Catania, collana estinta, torna ad essere fruibile dal pubblico italiano, per volontà del marito della Panvini-Carciotto, e grazie al web si spera di raggiungere il maggior numero possibile di utenti, andando oltre le logiche di protezione della conoscenza con l'utilizzo troppo rigido del diritto della proprietà intellettuale [8] che non aiuta la diffusione della cultura. Gerberto che *allestiva assiduamente una biblioteca* aveva *comprato con molto denaro [opere di] scrittori e copie di autori, aiutato dalla benevolenza e dalla cura degli amici della provincia* (ecclesiastica di Reims). Nella medesima lettera chiedeva ancora: *Indicheremo alla fine della lettera le opere che vorremmo che fossero copiate. A vostro comando spediremo la pergamena e le spese necessarie [per i copisti]* [2, Lettera 44]. Dunque l'Università Pontificia Regina Apostolorum cura, senza fini di lucro, l'edizione cartacea, coperte le spese dell'odierna pergamena di cellulosa; parimenti il testo è disponibile gratuitamente sul web sul sito *mirror* dell'IRSOL, Istituto Ricerche Solari di Locarno

http://irsol.ch/costantino_sigismondi/GerbertoEpistolario.pdf

In latino, per approfondire l'opera di Gerberto, esiste il vol. 139 della *Patrologia Latina*, che sul web è disponibile su http://la.wikisource.org/wiki/Patrologia_Latina_Vol_139_Silvester_II

Il testo Latino è da usarsi non soltanto come fonte di "versioni" per impraticchirsi sulle regole sintattiche e grammaticali, ma soprattutto per apprendere il gusto della traduzione come strumento di ricerca ed attualizzazione di conoscenze altrimenti irreperibili.

Basti pensare che ancora tutta la *Geometria* di Gerberto è da tradurre in Italiano, e solo l'*incipt* esiste tradotto in Francese. Come le opere di Gerberto, moltissima scienza medievale è ancora nascosta nella *Patrologia Latina*. La traduzione è l'unica maniera per renderla accessibile, ed una buona traduzione potrebbe essere frutto della collaborazione interdisciplinare di docenti di tutte le materie.

**Gerberto, esempio magistero e storiografia**

Altro aspetto da considerare è la sua probità, il suo rigore e morale e santità di vita. Nelle lettere emerge un continuo riferimento al Vangelo, al fare quaresime, digiuni e preghiere...ad avere il Cristo come modello: un Cristo comprensibile anche dai "gentili", viste le frequenti citazioni di classici latini. Non si tratta di un aspetto secondario della sua spiritualità e mostra la sua vocazione al dialogo interculturale, proiettato un millennio oltre i suoi tempi, ed iniziata durante il suo soggiorno in Catalogna con i primi contatti con la scienza araba.

La ricognizione della sua tomba è avvenuta nel 1648, ed il suo corpo, trovato intatto, si è dissolto al contatto con l'aria, ed ora resta solo l'epitaffio tombale murato dietro al pilastro di S. Filippo nella navata destra della Cattedrale del Laterano, dove fu sepolto per sua volontà. Egli riportò in auge la Cattedra del Laterano restaurandone idealmente il prestigio e la funzione simbolica di *Mater omnium Ecclesiarum* in un periodo storico così pieno di traversie per la chiesa di Roma. In questa veste fu riconosciuto e celebrato da papa Sergio IV, che, solo tra i suoi immediati successori, volle dedicargli l'epitaffio (antecedente al 1012) che ancora oggi possiamo leggere.

Successivamente la fama della sua scienza e l'invidia della sua carriera così straordinaria lo videro oggetto di leggende che parlano anche di magia, che sono figlie del loro tempo e di un'inevitabile distanza tra il genio di Gerberto e l'ordinarietà della gente comune. Nell'epoca della riforma nelle *Centurie* di Magdeburgo si sosterrà la tesi di simonia, grazie alla quale Gerberto sarebbe giunto alla carica pontificia, allo scopo di dimostrare la non continuità della successione apostolica sul soglio di Pietro. Una riabilitazione di Gerberto cominciò a farla nel 1629 il domenicano polacco Abraham Bzovsky (1563-1637) [9]. Dopo la pubblicazione dell'epistolario è cominciata con il Duchesne, proseguita dal Migne, dall'Olleris, e da Bubnov (tra la fine del XIX e l'inizio del XX secolo) che ha avuto il grande merito di cominciare ad inquadrare Gerberto nella storia della scienza nella posizione chiave che gli spetta.

**Conclusioni**

Oggi l'opera scientifica di Gerberto ci testimonia l'attività di un genio capace di fare sintesi di culture differenti come quella araba (per l'introduzione in Europa dell'astrolabio, del monocordo nella didattica musicale e le operazioni con le cifre indo-arabiche) e quella latina fondata sul pitagorismo ripreso da Boezio e mediato dalla manualistica alto-medievale. Il ruolo di Gerberto nella definizione del *curriculum studiorum* delle scuole cattedrali è stato determinante per le discipline tecnico-scientifiche, e la struttura dei suoi trattati quando venivano "rogati a pluribus" o quando rispondevano brevemente a singole questioni si presenta del tutto omogenea all'impostazione dialettica di opere scientifiche anche molto posteriori, ma sempre appartenenti all'ecumène latina. Ad esempio, pure nel *De Revolutionibus* di Copernico viene cercato il sostegno di Virgilio e dei classici latini come fondamento delle nuove teorie sul moto dei cieli.

Silvestro II, *di venerata memoria*, è stato ricordato dal papa Benedetto XVI [10] proprio al principio dell'anno dell'astronomia, e quando questo volge a conclusione esce in nuova edizione italiana il suo Epistolario.

Dagli *scriptoria* di Bobbio ai *server* su *world wide web* è trascorso un intero millennio, e dopo mille anni possiamo tornare ad onorare la memoria di Gerberto, studiando e facendo conoscere le sue opere, affinché si possa continuare a scegliere *il sicuro ozio degli studi invece dell'incerto affare delle guerre.*[2, Lettera 45].

Anche il lavoro di Maria Giulia Panvini Carciotto raggiungerà meglio il suo scopo, riaccendendo l'interesse verso una figura di enorme importanza sia nella Storia della Scienza che della Chiesa.

Dopo 1000 anni la riabilitazione di Gerberto è ad un livello mai raggiunto prima d'ora, ed è determinante in questo senso l'opera che studiosi e scienziati possono fare, anche per aiutare la stessa Chiesa a comprendere e a valorizzare una figura così poliedrica.

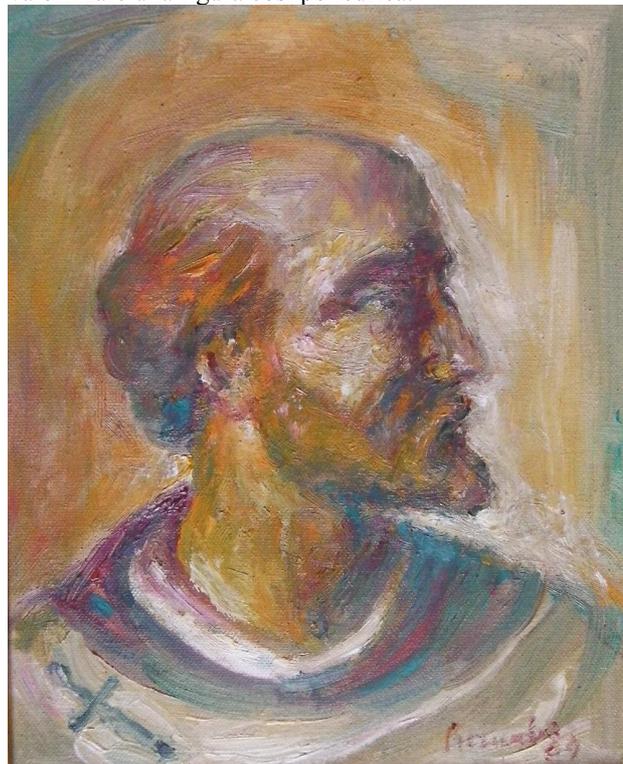

Maria Teresa Bernabei, Gerberto, olio su tela [*d'après*] (2009).